\documentclass[12pt]{article}

\usepackage{epsf}
\usepackage[dvips]{graphicx}

\begin{document}

\begin{center}
{\bfseries STUDY OF MULTIPARTICLE PRODUCTION BY GLUON DOMINANCE
MODEL (Part I)} \footnote {Talk given at the XVII International
Baldin Seminar "Relativistic Nuclear Physics and Quantum
Chromodynamics". JINR. September 27 to October 2, 2004, Dubna,
Russia. }

\vskip 5mm

 E.S.~Kokoulina$^{1,2 \dag}$, V.A.~Nikitin$^2$
 \vskip 5mm
 {\small (1) {\it GSTU, Belarus }, (2) {\it
JINR },
\\
$\dag$ {\it E-mail: Elena.Kokoulina@sunse.jinr.ru }}
\end{center}
\vskip 5mm

\begin{center}
\begin{minipage}{150mm}
\centerline{\bf Abstract}

\quad The gluon dominance model offers a description of
multiparticle production in $e^+e^-$ - annihilation and
proton-proton collisions. The multiplicity distributions in
$e^+e^-$ annihilation are well described. The energy dependence of
model parameters gives the dynamic parton stage and hadronisation
picture. It is shown that this model has confirmed oscillations in
sign of the ratio of factorial cumulant moments over factorial
moments of the increasing order.

\quad The collective behavior of secondary particles in
$pp$-interactions at 70 GeV/c is studied  in the project {\bf
"Thermalization"}. An active role of gluons is shown in the
multiparticle dynamics. This paper gives a simple thermodynamic
interpretation of interactions mentioned above.
\end{minipage}
\end{center}

\vskip 5mm

\section{Introduction}

At present to investigate and construct a contemporary picture of
nuclear matter structure requires to develop new methods and
approaches.

Since 80's the Quark-Gluon Plasma conception has undergone a lot
of changes after the experiments carried out at CERN (SPS) and BNL
(RHIC). Different approaches are used to explain extraordinary
phenomena in behavior of the new matter produced at high energy of
nuclear collisions  \cite{WhPa}.

Still there is no single theory nor the model that could explain
all the results obtained at RHIC and SPS. We would like to find a
solution for this difficult problem by using multiparticle
production (MP) in hadron and nucleus interactions. Up to now the
nature of "soft" hadronic events has not been fully understood.

A new way to investigate MP at high energy is offered in this work
by means of construction a model based on the multiplicity
distribution (MD) description using the QCD and phenomenological
scheme of hadronisation. The model description of MD in $e^+e^-$
annihilation is given in section 2. The application of this model
approach to pp-interaction can be found in section 3.

\section{MD for $e^+e^-$-annihilation at high energies}

The $e^+e^-$ annihilation is one of the most suitable to study MP.
It can be realized through the formation of virtual $\gamma$ or
$Z^0$--boson which then decays into two quarks:
\begin{equation}
\label{1} e^+e^-\rightarrow(Z^0/\gamma)\rightarrow q\bar q.
\end{equation}
The $e^+e^-$--reaction is simple for analysis, as the produced
state is pure $q\overline q$. It is usually difficult to determine
the quark species on the event-by-event basis. The experimental
results are averaged over the quark type. Perturbative QCD (pQCD)
may be applied to describe the process of parton fission (quarks
and gluons) at big virtuality, because strong coupling $\alpha_s$
is small. This stage can be called as the stage of cascade. When
partons get small virtuality, they change into hadrons, which we
observe. At this stage we can not apply pQCD. Therefore
phenomenological models are used to describe hadronisation in this
case.

Parton spectra the quark and gluon fission in QCD were studied by
K.~Konishi, A.~Ukawa and G.~Veneciano. The probabilistic nature of
the problem has been established \cite{KUV} while working at the
leading logarithm approximation and avoiding IR divergences by
considering finite $x$. Studying MP at high energy we used ideas
of A.~Giovannini~\cite{GIO} to describe  quark-gluon jets as
Markov branching processes. Giovannini proposed to interpret the
natural QCD evolution parameter $Y=\frac{1}{2\pi
b}\ln[1+\alpha_{s} b\ln(\frac{Q^2}{\mu^2})]$ , where $2\pi
b=\frac{1}{6}(11N_C-2N_f)$ for a theory with $N_C$ colours, $N_f$
flavours and virtuality $Q$ as the thickness of the jets and their
development as the Markov process.

Three elementary processes contribute into QCD jets: (1) gluon
fission; (2) quark bremsstrahlung and (3) quark pair production.
Let $A\Delta Y$ be the probability that gluon will convert into
two gluons in the infinitesimal interval $\Delta Y$,
$\widetilde{A}\Delta Y$ be the probability that  quark will
radiate a gluon, and $B\Delta Y$ be the probability that a
quark-antiquark pair will be produced from a gluon. A.Giovannini
constructed a system of differential equations for generating
functions (GF) of quark $Q^{(q)}$ and gluon $Q^{(g)}$ jets
$$
\frac{dQ^{(q)}}{dY}=\tilde{A}Q^{(q)}(Q^{(g)}-1), \quad
\frac{dQ^{(g)}}{dY}=A(Q^{(g)2}-Q^{(g)})+B(Q^{(q)2}-Q^{(g)})
$$
and obtained explicit solutions of MD in the case $B=0$ (process
of quark pair production is absent)
\begin{equation}
\label{4} P_{0}^P(Y)=\left (\frac{k_p}{k_p+\overline m}\right
)^{k_p}, P_{m}^P(Y)=\frac{k_p(k_p+1)\dots(k_p+
m-1)}{m!}\left(\frac{\overline m} {\overline
m+k_p}\right)^{m}\left( \frac{k_p}{k_p+\overline m}\right)^{k_p},
\end{equation}
where $k_p=\widetilde A/A$, $\overline m=k_p (e^{AY}-1)$ is the
mean gluon multiplicity. These MD are known as negative binomial
distribution (NBD). The GF for them is
\begin{equation}
\label{6} Q^{(q)}(z,Y)=\sum\limits_{m=0}^{\infty} z^{m} P_{m}(Y)=
 \left[1+ \overline m/k_p (1-z)\right]^{-k_p}.
\end{equation}

Two Stage Model (TSM) \cite{TSM} was taken (\ref{4}) to describe
the cascade stage and added with a sub narrow binomial
distribution for the hadronisation stage. We have chosen it basing
on the analysis of experimental data in $e^+e^-$- annihilation
lower than 9 GeV. Second correlation moments were negative at this
energy. The choice of such distributions was the only one that
could describe the experiment. We suppose  that the hypothesis of
soft decoloration is right. Therefore we add the hadronisation
stage to the parton stage for the sake of its factorization. MD in
this process can be written as follows:
\begin{equation}
\label{7} P_n(s)=\sum\limits_mP^P_mP_n^H(m,s),
\end{equation}
where $P_m^P$ is MD for partons (\ref{4}), $P_n^H(m,s)$ - MD for
hadrons produced from $m$ partons at the stage of hadronisation.
Further we substitute variable $Y$ on a center of masses energy
$\sqrt s$. MD of hadrons $P_n^H$ formed from one parton and their
GF $Q^H_p(z)$ are \cite{TSM}
\begin{equation}
\label{12} P_n^H=C^n_{N_p}\left(\frac{\overline n^h_p}
{N_p}\right)^n\left(1-\frac{\overline n_p^h}
{N_p}\right)^{N_p-n},\quad Q^H_p=\left[1+\frac{\overline n^h_p}
{N_p}(z-1)\right]^{N_p},
\end{equation}
where $C_{N_p}^n$ - binomial coefficient, $\overline n^h_p$ and
$N_p$ ($p=q,g$) have a sense of mean multiplicity and maximum of
secondary hadrons are formed from parton at the stage of
hadronisation.

 MD of hadrons in $e^+e^-$ annihilation are
determined by convolution of two stages (cascade and
hadronisation)
\begin{equation}
\label{13} P_n(s)=\sum\limits_{m=o}^{\infty}
P_m^P\frac{1}{n!}\frac{\partial^n}{\partial z^n}
(Q^H)^{2+m}|_{z=0},
\end{equation}
where $2+m$ is the total number of partons (two quarks and $m$
gluons).

We introduce parameter $\alpha=N_g/N_q$ to distinguish the
hadrons, produced from quark or gluon. Also we have carried out
simplification for designation $N=N_q$, $\overline n^h=\overline
n^h_q$. Introducing expressions (\ref{4}), (\ref{12}) in
(\ref{13}) and differentiating on $z$, we obtain MD of hadrons in
the process of $e^+e^-$ annihilation in TSM
\begin{equation}
\label{15} P_n(s)= \sum\limits_{m=0} ^{M_g}P_m^PC^n_{(2+\alpha
m)N} \left(\frac{\overline n^h} {N}\right)^n\left(1-\frac
{\overline n^h}{N}\right)^{(2 +\alpha m)N-n}.
\end{equation}

The results of comparison of expression (\ref{15}) with
experimental data \cite{DAT} are shown in Figs. 1-2. We have
obtained that MD in TSM (solid curve) describe well the
experimental $e^+e^-$-data from $14$ to $189$ GeV \cite{NPCS}. The
mean gluon multiplicity $\overline m$ has a tendency to rise, but
lower than the logarithmic curve. Values $k_p$ remain $\sim 10$ at
almost all energies. One of the most interesting physical senses
of this parameter is temperature $T$\cite{TEM}: $T\sim k_p^{-1}$.

The next picture of the hadronisation stage is discovered in
conformity with parameters of the second stage: $N$, $\overline
n^h$ and $\alpha$ . The first parameter $N$ determines the maximum
number of hadrons, which can be formed from quark while its
passing through this stage. We can imagine that fission is
continuous but process (3) (formation $q\bar q$~pair) becomes
comparable with the other ones (1),(2). We can not reveal a steady
energy rise or fall for $N$.

The second parameter $\overline n^h$ has a sense of the mean
hadron multiplicity from quark at the second stage. We have found
out the tendency to a weak rise with big scattering. The value of
$\overline n^h$ is about $5-6$ in the research region. A possible
explanation of these rocks for $N$ and $\overline n^h$: only two
initial quarks exist among a lot of newly born gluons at the
beginning of hadronisation.

The last parameter $\alpha$ was introduced to compare the quark
and gluon hadronisation. It is equal to $0.2$ with some
deviations. If we know $\alpha$, then we can determine $N_g=\alpha
N$ and $\overline n_g^h= \alpha \overline n^h$ for gluon. It is
surprising that gluon parameters remain constant without
considerable deviations in spite of the indirect finding: $N_g\sim
3-4$ and $\overline n_g^h\sim 1$ (Fig. 3-4). Therefore we can
confirm the universality of gluon hadronisation. The fact that
$\alpha <1$ shows that hadronisation of gluons is softer than that
of quarks.

It was shown \cite{OSC} that the ratio of factorial cumulative
moments $K_q$ over factorial moments $F_q$ changes the sign as a
function of the order. We use MD formed in TSM to explain this
phenomenon. $F_q$ and $K_q$ are obtained from the relations
\begin{equation}
\label{17} F_q=\sum\limits_{n=q}^{\infty} n(n-1)\dots(n-q+1)P_n,
\quad K_q=F_q-\sum\limits_{i=1}^{q-1}C_{q-i}^{i} K_{q-i}F_i .
\end{equation}
The ratio of their quantities is $H_q=K_q/F_q$. The generating
function for MD of hadrons (\ref{15}) in $e^+e^-$ annihilation
Q(z) is the convolution
\begin{equation}
\label{20} Q(z)=\sum\limits_{m=0}P_m^g[Q_g^H(z)]^m Q^2_q(z)=
Q^g(Q^H_g(z))Q^2_q(z).
\end{equation}
We calculate $F_q$ and $K_q$ in TSM, by using (\ref{20}) and the
sought-for expression for $H_q$ will be \cite{NPCS}
\begin{equation}
\label{26} H_q=\frac{\sum\limits_{m=1}k_p\alpha m (\alpha
m-\frac{1}{N})\dots(\alpha m-\frac{q-1} {N})(\frac{\overline
m}{\overline m+k_p})^m\frac{1}{m}- 2(-1)^q\frac{(q-1)!}{N^{q-1}}}
{\sum\limits_{m=0}(2+\alpha m)(2+\alpha m-
\frac{1}{N})\dots(2+\alpha m-\frac{q-1}{N})P_m}.
\end{equation}
The comparison with experimental data \cite{OSC} has shown that
(\ref{26}) describes the ratio of factorial moments (Fig. 5). The
minimum is seen at $q=5$. We have obtained that in the region
before $Z^0$ (91.4 GeV), $H_q$ oscillates in the sign only with
the period equal to $2$ and changes the sign with parity $q$. At
higher energies the period is increased to $4$. It can be
explained by influence of a more developed cascade of partons with
narrow hadronisation.

\section{MD in $pp$-interactions}

The study of MD in $pp$ interactions is implemented in the
framework of the project "Thermalization". This project is aimed
at studying the collective behavior of secondary particles in
proton-proton interactions at 70 GeV/c \cite{THE}. On the basis of
the present understanding of hadron physics, protons consist of
quarks and gluons. After the inelastic collision the part of the
energy of the initial motive protons are transformed to the inside
energy. Several quarks and gluons become free. Our model study has
shown that quark branching of initial protons in pp interactions
is almost absent from 70 to 800 GeV/c. MP are realized by active
gluons. Domination of gluons was first proposed by S. ~Pokorski
and L.~Van ~Hove \cite{PVH}.

Our choice of the MP model is based on comparison with the
experimental partial cross section $\sigma (n_{ch})$ in $pp$
interaction at 70 GeV/c on the U-70 accelerator~\cite{BAB} and the
present picture of strong interactions.

At the beginning of 90s a successful description of MD at this
energy was realized by the quark model (Fig. 6) \cite{CHI}. This
model suggests that one proton quark pair, two pairs or three can
collide and fragment into hadron jets. MD in quark jets were
described by Poisson. Second correlation moments of charged
particles for MD in this model will be always negative. It is
known they are become positive at higher energies. In this model
gluons are absent. The calculation by the MC PHYTHIA code has
shown that the standard generator predicts a value of the cross
section which is in a reasonably good agreement with the
experimental data at small multiplicity ($n_{ch}<10$) but it
underestimates the value $\sigma(n_{ch})$ by two orders of the
magnitude at $n_{ch}=18$ (Fig. 6).

We have managed to build a scheme of hadron interactions to
describe MD with the quark-gluon language as well as to
investigate the high multiplicity region. The mentioned models are
very much sensitive in this region.

We consider that at the early stage of $pp$ interactions the
initial quarks and gluons take part in the formation of
quark-gluon system (QGS). They can give branches. We offer two
model schemes. In the first scheme we study hadroproduction with
account of the parton fission inside the QGS and build the two
stage model with branch (TSMB). If we are not interested in what
is going inside QGS, we come to the thermodynamical model (TSTM).
Onward we name models involving active gluons into hadroproduction
as the gluon dominance models (GDM) \cite{MGD}.

We begin our MD analysis with the scheme of branch. MD for quark
and gluon jets may be described NBD and Farry distributions
\cite{GIO}, accordingly. On the hadronisation stage we have taken
a binomial distribution (\ref{12}). As in TSM we have used a
hypothesis of soft decoloration for quarks and gluons at their
while passing of this stage and add the hadronisation stage to the
branch one by means of factorization
\begin{equation}
\label{27} P_n(s)=\sum\limits_{m} P_m^P(s)P_n^H(m),
\end{equation}
where $P_n(s)$ - resulting MD of hadrons, $P_m^P$ - MD of partons
(quarks and gluons), $P_n^H(m)$~- MD of hadrons (second stage)
from $m$ partons. Generating function (GF) for MD in hadron
interactions is determined by convolution of two stages:

\begin{equation}
\label{28} Q(s,z)=\sum\limits_{m}P_m^P(s)\left(Q^H(z)\right)^m=
Q^P(s,Q^H(z)),
\end{equation}
where $Q^H$ and $Q^P$ are GF for MD at hadronisation stage and in
QGS.

At the beginning of research we took model where some of quarks
and gluons from protons participate in the production of hadrons.
Parameters of that model had values which differed very much from
parameters obtained in $e^+e^-$- annihilation, especially
hadronisation parameters. It was one of the main reasons to refuse
the scheme with active quarks. After that we chose the model where
quarks of protons did not take part in the production of hadrons,
but remained inside of the leading particles. All of the newly
born hadrons were formed by gluons. We name these gluons active.
They could give a branch before hadronisation.

It is important to know how much active gluons are into QGS at the
first time after the impact of protons. We can assume that their
number may change from zero and higher. It is analogous with the
impact parameter for nucleus. Only in the case of elastic
scattering the active gluons are absent. The simplest MD to
describe the active gluons formed in the moment of impact is the
Poisson distribution $P_k= e^{-\overline k} \overline k^k/{k!}$,
where $k$ and $\overline k$ are the number and mean multiplicities
of active gluons, correspondingly.

To describe MD in the gluon cascade formed by the branch process
of $k$ active gluons, we have used the Farry distribution
\cite{GIO}
\begin{equation}
\label{30} P_m^B(s)= \frac{1}{\overline
m^k}\left(1-\frac{1}{\overline m} \right)^{m-k}\cdot \frac
{(m-1)(m-2)\cdots (m-k+1)} {(k-1)!},
\end{equation}
\begin{equation}
\label{31} P_m^B(s)=\frac {1}{\overline m }\left (1-\frac
{1}{\overline m}\right)^{m-1},
\end{equation}
at $k>1$ - (\ref{30})  and at $k=1$ - (\ref{31}). m and $\overline
m$ are the number of secondary gluons and mean multiplicites of
them (averaged to all gluons). Expressions (\ref{30})-(\ref{31})
have been obtained from the assumption about the independent
branch of gluons
\begin{equation}
\label{33} Q_k^B= (Q_1^B)^k =\frac{z^k}{\overline m^k}\left[1-z
\left(1-\frac{1}{\overline m} \right)\right]^{-k},\quad Q_1^B=
\frac{z}{\overline m}\left[1-z \left(1-\frac{1}{\overline m}
\right)\right]^{-1}.
\end{equation}
In the case $k=0$ (the impact was elastic and active gluons are
absent) MD of hadrons in pp-scattering is equal to
$P_2^{el}(s)=e^{-\overline k}$.

On the second stage some of active gluons may leave QGS and
transform to real hadrons. We can name that gluons "evaporated".
Let us introduce parameter $\delta$ as the ratio of evaporated
gluons, leaving QGS, to all active gluons, which may transform to
hadrons. Our binomial distributions for MD of hadrons from the
evaporated gluons on the stage of hadronisation are

\begin{equation}
\label{34} P_n^H(m)= C^{n-2}_{\delta  mN}\left(\frac{\overline
n^h} {N}\right)^{n-2}\left(1-\frac{\overline n^h}
{N}\right)^{\delta mN-(n-2)}.
\end{equation}
In this expression the gluon parameters are $\overline n^h$ and
$N$ (without index "g") which have the same meaning that of the
quark parameters. An effect of two leading protons is also taken
into account. GF for MD (\ref{34}) has the following form:

\begin{equation}
\label{35} Q_m^H= \left (Q_1^H \right )^{\delta m}= \left
[1-\frac{\overline n^h}{N}\left ( 1-z \right ) \right ]^{\delta
mN}, \quad Q^H_1=\left [1-\frac{\overline n^h}{N}(1-z) \right
]^{N}.
\end{equation}

MD of hadrons in the process of proton-proton scattering in two
stage gluon model with branch (TSMB) is

$$
P_n(s)= \sum\limits_{k=1}^{MK}\frac{e^ {-\overline k} \overline
k^k}{k!}\sum\limits_{m=k}^{MG} \frac {1}{\overline m^k} \frac
{(m-1)(m-2)\dots(m-k+1)}{(k-1)!} \cdot
$$

\begin{equation}
\label{37} \cdot \left(1-\frac{1}{ \overline m}\right)^{m-k}
C^{n-2}_{\delta mN}\left(\frac{\overline n^h}
{N}\right)^{n-2}\left(1-\frac{\overline n^h} {N}\right)^{\delta
mN-(n-2)}.
\end{equation}

In comparison with experimental data \cite{BAB} the numbers of
gluons in sums on $k$ and $m$ were restricted by values $MK$ and
$MG$ as the maximal possible number of gluons on the transition.
For comparison we have taken the data at $69$ GeV/c because they
do not differ from data at $70$ GeV/c \cite{BAB}. $\chi^2/$ndf in
are equal to about $\sim 1/3$ at $70$ GeV/c and $\sim 5$ at $69$
GeV/c and the parameters are similar. We obtained $N=40$(fix),
$\overline m=2.61 \pm 0.08$, $\delta = 0.47 \pm 0.01$, $\overline
k=2.53 \pm 0.05$, $\overline n^h=2.50 \pm 0.29$ from the
comparison with \cite{BAB} . We can conclude that the branch
processes are absent, since parameters $\overline m$ and
$\overline k$ are equal to the errors. The fraction of the
evaporated gluons is equal to 0.47. A maximal possible number of
hadrons from the gluon looks very much like the number of partons
in the glob of cold quark-gluon plasma of L.Van Hove \cite{LVH}.
The gluon branch should be very active inside of QGS. At the fixed
parameter of hadronisation $\overline n^h$ equal to $1.63$ (see
below the thermodynamic model) the fraction of the evaporated
gluons $\delta$ is about 0.73. After the evaporation the part of
active gluons do not convert into hadrons. They stay in QGS and
become sources of soft photons (SP). Further we will analyze the
experimental effect of SP excess (it is impossible to describe
them by means QED).
\begin{figure}
\begin{minipage}[b]{.3\linewidth}
\centering
\includegraphics[width=\linewidth, height=2in, angle=0]{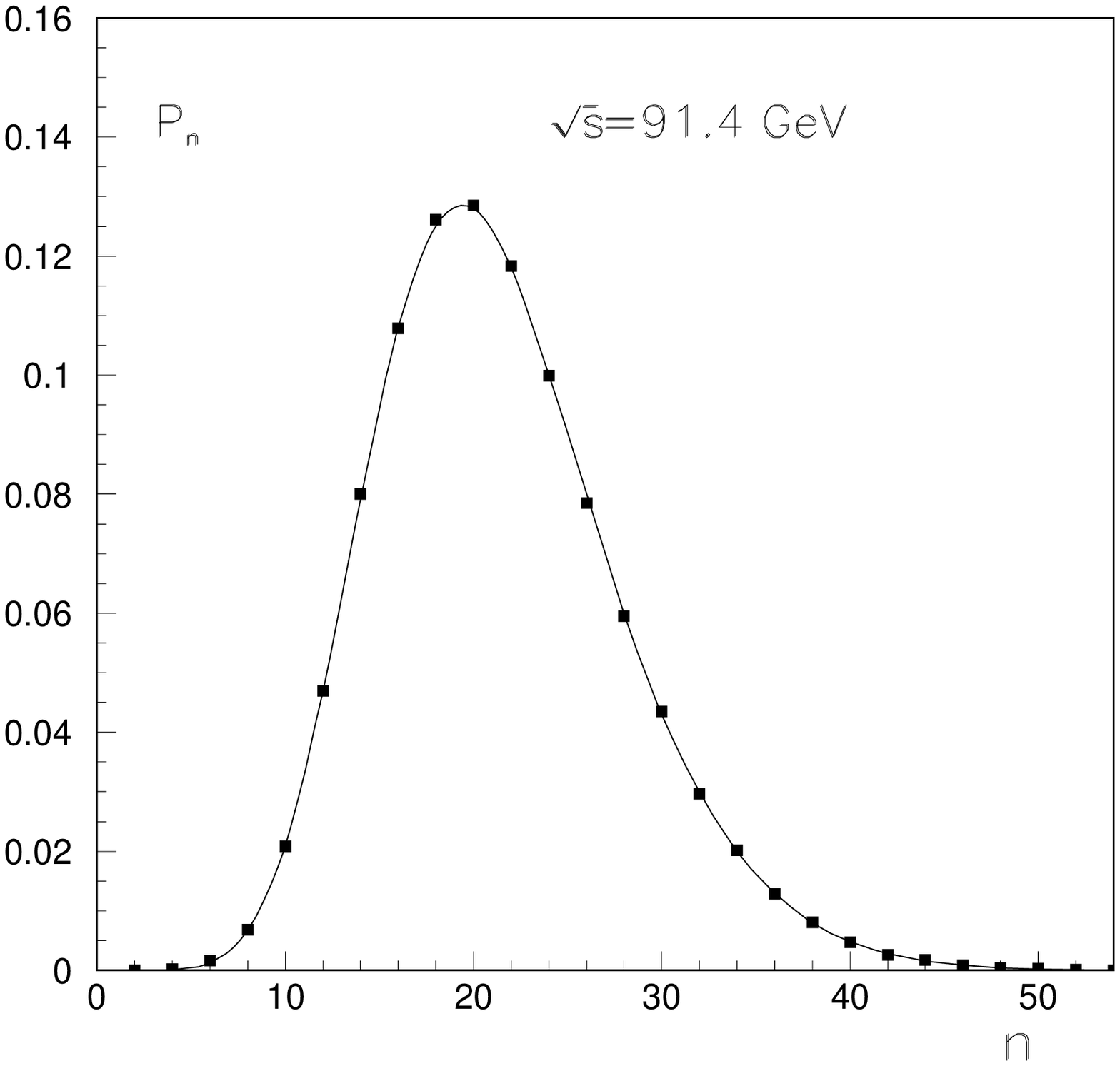}
\caption{MD at 91.4 GeV.} \label{1dfig}
\end{minipage}\hfill
\begin{minipage}[b]{.3\linewidth}
\centering
\includegraphics[width=\linewidth, height=2in, angle=0]{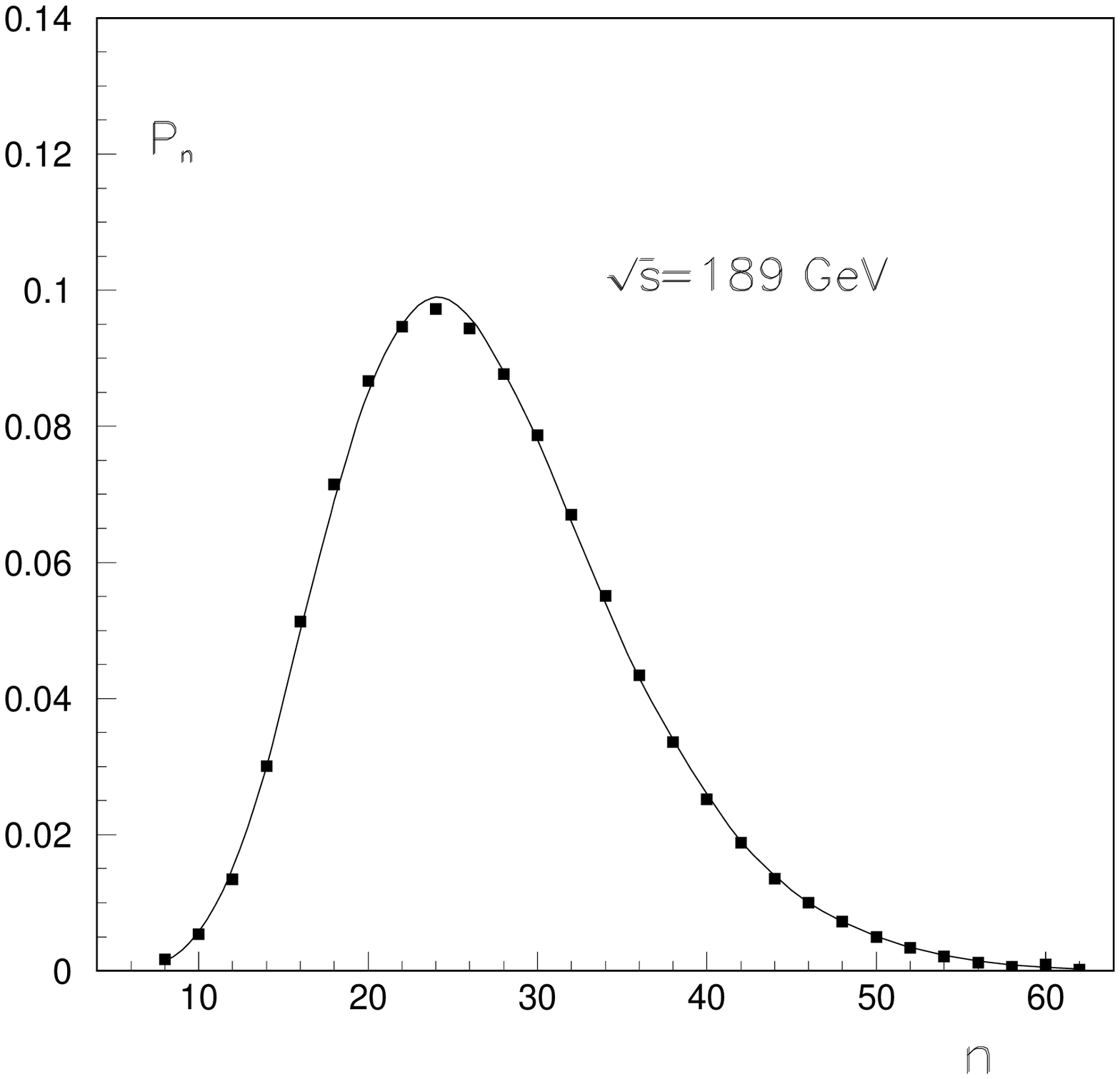}
\caption{MD at 189 GeV.} \label{2dfig}
\end{minipage}\hfill
\begin{minipage}[b]{.3\linewidth}
\centering
\includegraphics[width=\linewidth, height=2in, angle=0]{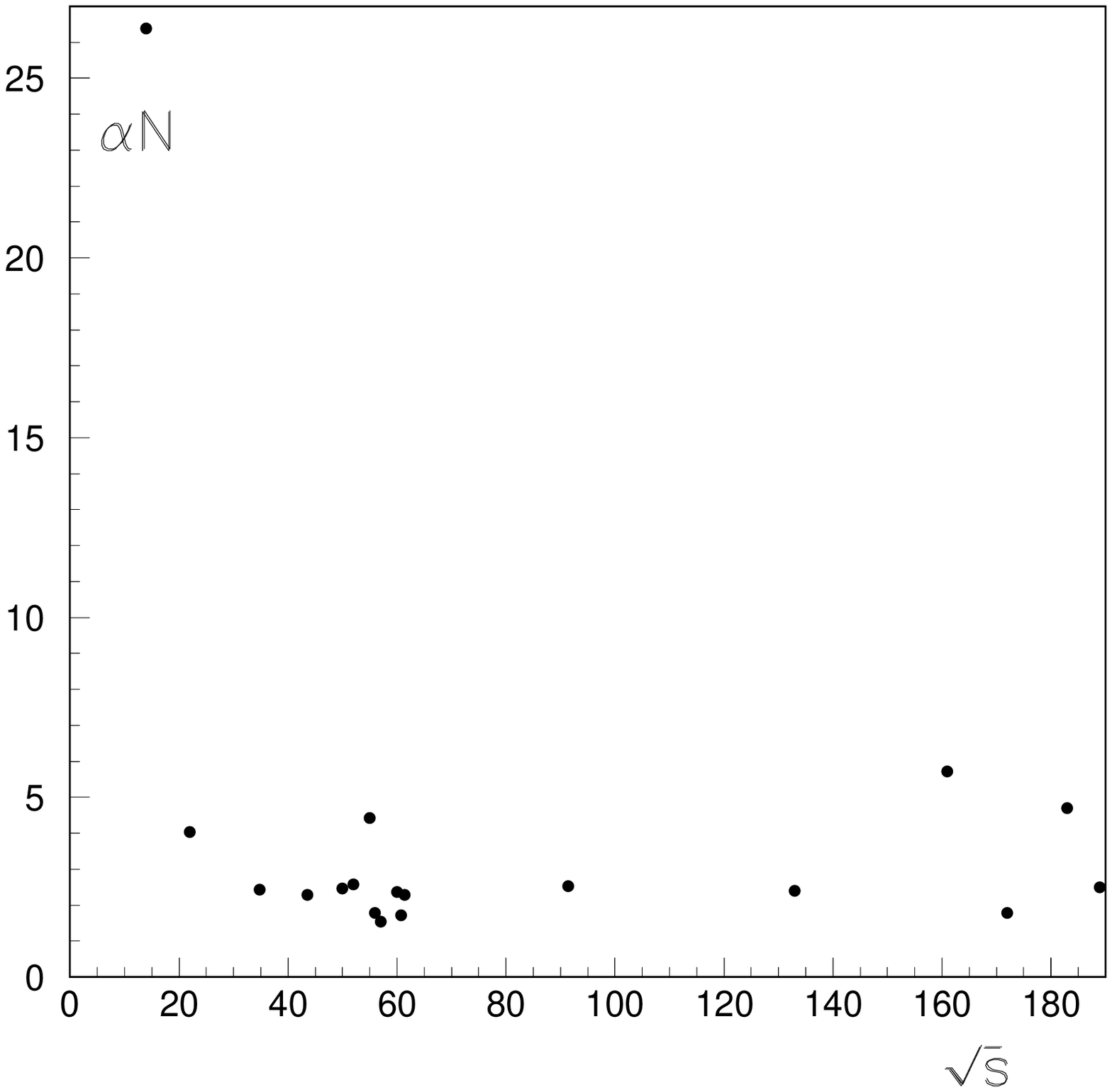}
\caption{$N_g=\alpha N_q$.} \label{3dfig}
\end{minipage}
\end{figure}

In the thermodynamic model without branches the active gluons
which appear in the moment of the impact may leave QGS and
fragment to hadrons. We consider that active gluons evaporated
from QGS have Poisson MD with a mean multiplicity $\overline m$.
Using the idea of the convolution of two stages (\ref{27}) as well
as the binomial distribution for hadrons from gluons we obtain MD
of hadrons in pp-collisions in framework of the two stage
thermodynamic model (TSTM):
\begin{equation}
\label{38} P_n(s)=\sum\limits_{m=0}^{ME}\frac{e^{-\overline m}
\overline m^m}{m!} C^{n-2}_{mN}\left(\frac{\overline n^h}
{N}\right)^{n-2}\left(1-\frac{\overline n^h} {N}\right)^{mN-(n-2)}
(n>2),
\end{equation}
$P_2^{el}(s)=e^{-\overline m}$. Our comparison (\ref{38}) with the
experimental data \cite{BAB} (see Fig. 8) gives values of
parameters $N=4.24 \pm 0.13$, \quad $\overline m=2.48 \pm 0.20$,
\quad $\overline n^h=1.63\pm 0.12$, and the normalized factor
$\Omega =2$ with $\chi ^2/$ndf $\sim 1/2$. We are restricted in
sum~(\ref{38}) $ME=6$ (the maximal possible number of evaporated
gluons from QGS). The found gluon parameters $N$ and $\overline
n^h$ agree with the values of these parameters obtained at the
$e^+e^-$ annihilation \cite{NPCS}. From TSTM the maximal possible
number of charged particles is $26$. This quantity is the product
of maximal multiplicities of active gluons and of the maximal
number of hadrons forming from one gluon $ME \cdot N$. In TSMB
there are no restrictions of this sort.
\begin{figure}
\begin{minipage}[b]{.3\linewidth}
\centering
\includegraphics[width=\linewidth, height=2in, angle=0]{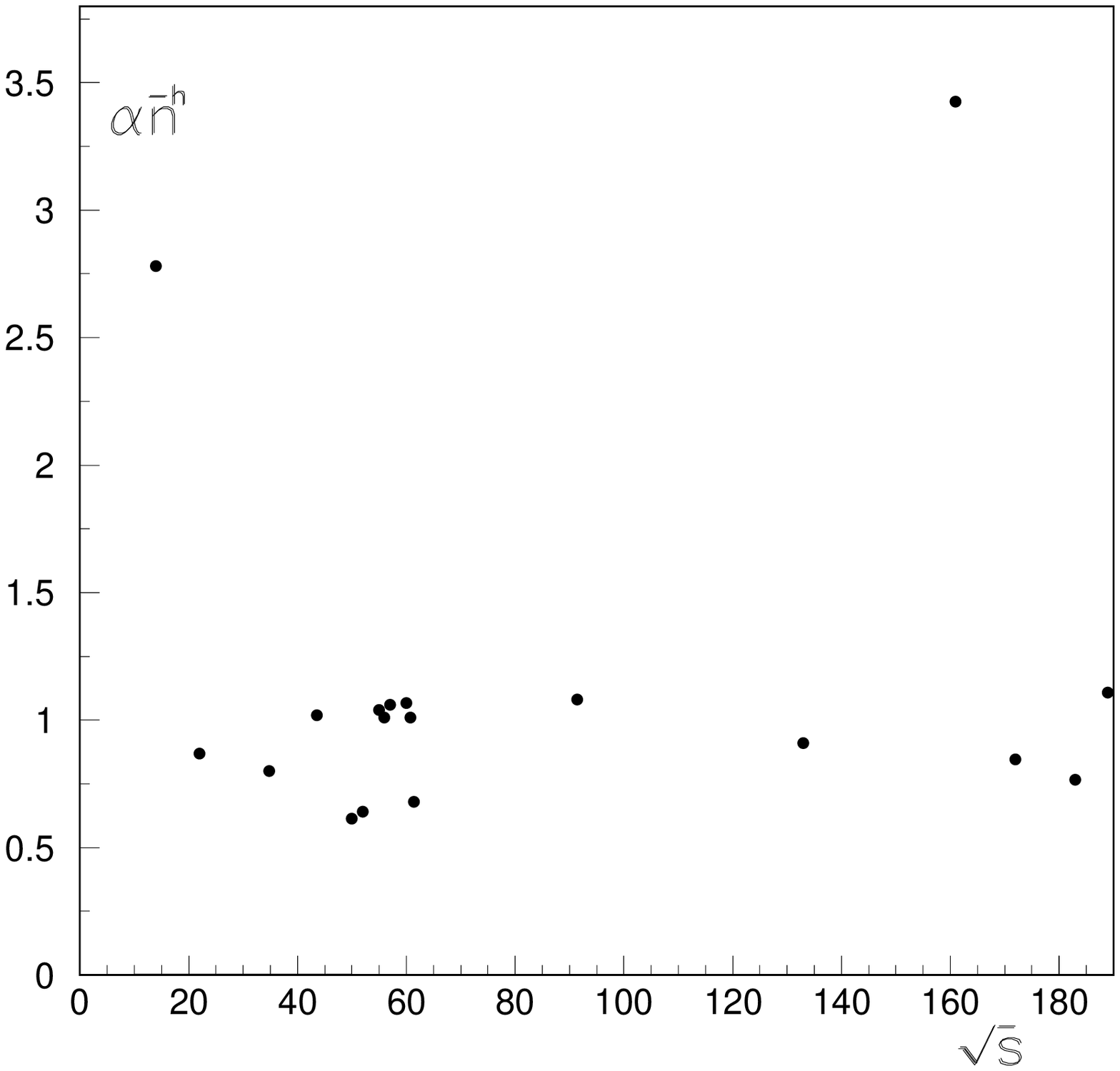}
\caption{$\overline n_g^h=\alpha \overline n_q^h$ .}
\label{13dfig}
\end{minipage}\hfill
\begin{minipage}[b]{.3\linewidth}
\centering
\includegraphics[width=\linewidth, height=2in, angle=0]{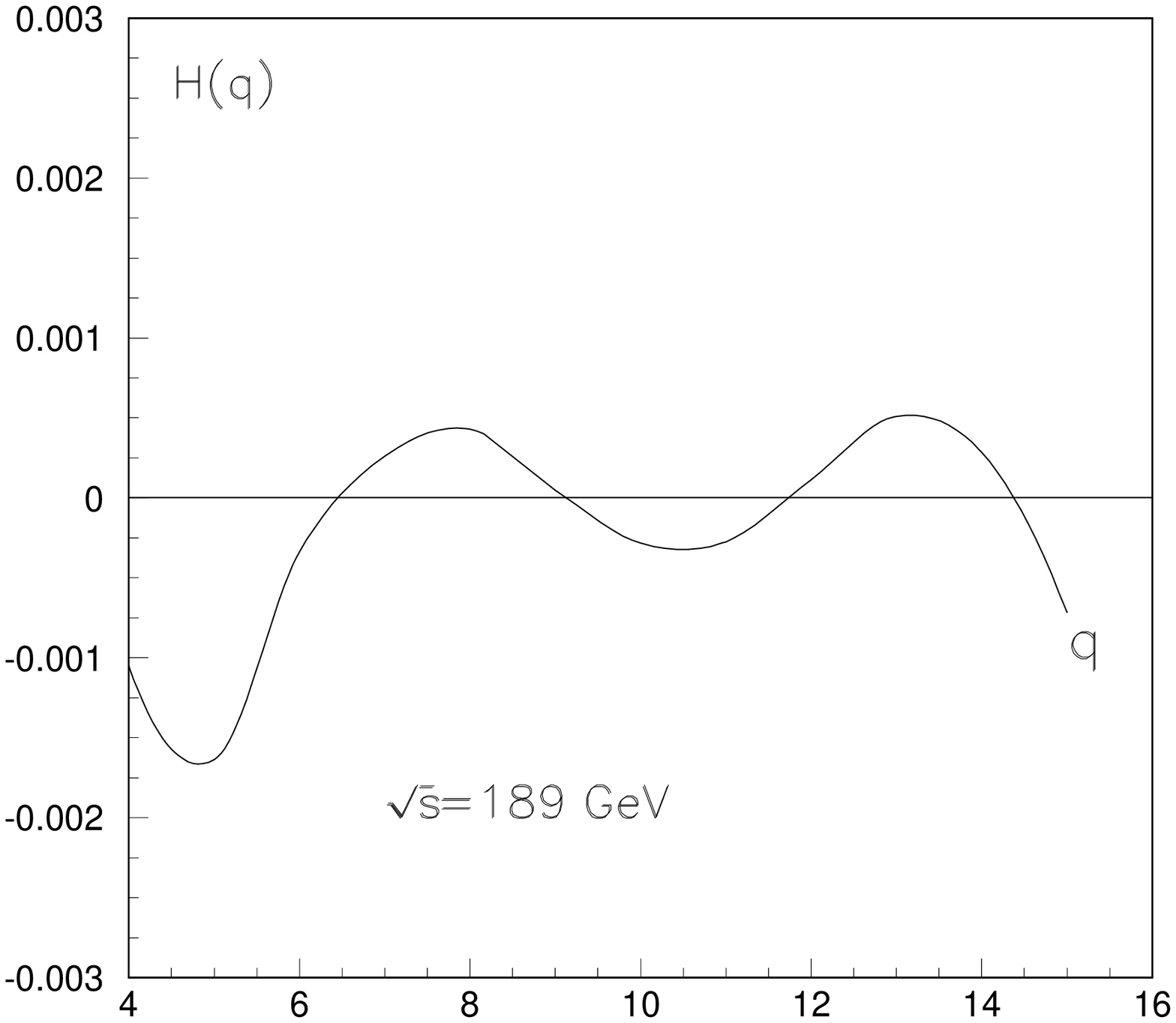}
\caption{$H_q$ at 189 GeV.} \label{14dfig}
\end{minipage}\hfill
\begin{minipage}[b]{.3\linewidth}
\centering
\includegraphics[width=\linewidth, height=2in, angle=0]{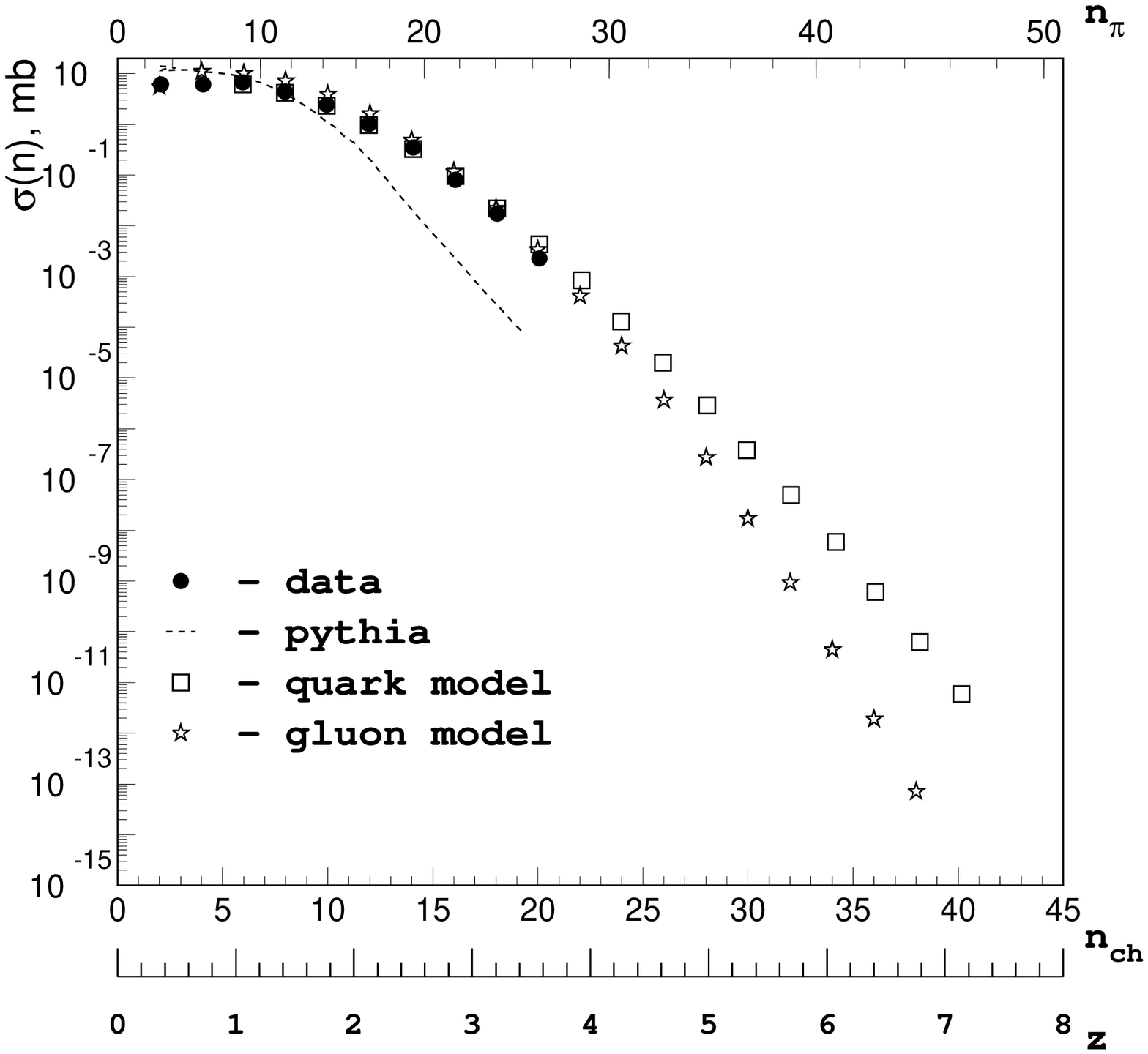}
\caption{$\sigma (n)$ in $pp$.} \label{15dfig}
\end{minipage}
\end{figure}

{\it Project "Thermalization" is partially supported by RFBR grant
03-02-16869}.

\end{document}